\documentclass{revtex4}
\usepackage{graphicx}
\usepackage{amsmath}
\usepackage{fancyhdr}
\begin{document}

\title{{\small{2005 International Linear Collider Workshop - Stanford,
U.S.A.}}\\ 
\vspace{12pt}
Pseudo-Axions in Little Higgs Models} 

%

\preprint{DESY 05--114}
\author{W.~Kilian, \underline{J.~Reuter}}
\affiliation{DESY, D-22603 Hamburg, Germany}
\author{D.~Rainwater}
\affiliation{Dep.~of Physics and Astronomy, University of Rochester, Rochester, NY 14627, USA}

\begin{abstract}
In Little Higgs models, the Higgs mass is stabilized at the one-loop
level by the mechanism of collective symmetry breaking. Typically, the
basic ingredient of these models is a large(r) global symmetry group
spontaneously broken at a scale of several TeV. The (light) Higgs 
appears as a Goldstone boson corresponding to a non-diagonal broken
generator of this global symmetry. There could also be physical
pseudoscalar particles present which belong to diagonal generators,
having the properties of axions with masses in the range from several
GeV up to the electroweak scale. We investigate the interesting
phenomenology of these pseudo-axions at the linear collider as well as
the photon collider.   

\end{abstract}

\maketitle

\thispagestyle{fancy}

\section{LITTLE HIGGS MODELS} 

A motivation for physics beyond the Standard Model (SM) in the
electroweak (EW) sector lies in the vast difference between the Planck (or
unification) and the EW scale (hierarchy problem), that requires some
stabilization of light scalar masses (fine tuning problem). In
contrast to the supersymmetric solution to that problem where the
quadratic sensitivity of the scalar masses to the cut-off is
cancelled above the SUSY breaking scale between partners of opposite
statistics, models have been constructed that contain a spontaneously
broken global symmetry with the Higgs being light because it is one
of the Goldstone bosons appearing in this breaking. In the simplest
variant this construction fails because the scale for new strong
interactions is too close to the electroweak scale, leaving traces in
the low-energy effective action which would have shown up at LEP and
Tevatron. There are two ways to evade these complications, either to use
non-simple global groups (deconstruction models) or to entangle the
global with the local symmetry breaking to forbid one-loop
contributions to the Higgs mass parameters \cite{little}. This shifts
the scale where new strong dynamics naturally appears by one order of
magnitude upwards. If the Higgs is among the Goldstone bosons of a
broken global symmetry group, one always has a reduction of the rank
of the global group. The Higgs thereby corresponds to a broken
non-diagonal generator like the kaon in chiral symmetry breaking. 


\section{PSEUDOSCALARS IN DIFFERENT MODELS}

In most Little Higgs models there is a
diagonal generator in the global symmetry breaking which 
corresponds to an (anomalous) unbroken global $U(1)$ subgroup. The
boson that parameterizes this subgroup is analogous to the
$\eta^{(\prime)}$ in chiral symmetry breaking, i.e. it couples like a
pseudoscalar to fermions. Hence, its properties resemble those of an
axion-like particle, so we call it the {\em pseudo-axion} of Little
Higgs models \cite{axions}. In order to lift the bounds (mainly from
astrophysics) on axions, we have to assume that the $U(1)$ symmetry is
explicitly broken and the relation between the mass of the axion and
its (loop-induced) coupling to photons is shifted. This is plausible,
since the Yukawa interactions of the non-linear extended Higgs-axion 
multiplet representation break the anomalous $U(1)$ anyway. 

In some cases, these additional (pseudo-)scalar
degrees of freedom have been absorbed as longitudinal modes of
heavy neutral vector bosons ($Z'$), which are quite easy
to detect either at LHC or at ILC. We consider the case where the
additional group remains ungauged (and can therefore be anomalous), so
that the pseudoscalar is part of the physical spectrum.  

Particles analogous to the Little Higgs pseudo-axion exist in other
models of EWSB like technicolor, topcolor and the NMSSM (for an
overview and a list of references, cf.~\cite{pdg}). We focus on the
influence of this particle on the EW observables and on the
phenomenology of the heavy quark states present in Little Higgs
models. 

Generically, the $\eta$ is an EW singlet which parameterizes the
anomalous $U(1)$ subgroup as $\xi = \exp[i\eta/F]$, where $F$ is the
vacuum expectation value of the global symmetry breaking, assumed to be
of the order of $1-5$ TeV. If it were an exact Goldstone boson, there
would be no potential for the $\eta$, so it would be exactly massless.
This is ruled out by the non-existence of long-range forces. As
already mentioned the fermion couplings (Yukawa interactions) break
the global symmetry explicitly, generating a potential and a mass for
the $\eta$, so that the astrophysical axion bounds are evaded. In order
not to reintroduce a hierarchy problem, power-counting demands a
mass $m_\eta \lesssim v \sim 250\,{\rm GeV}$. All couplings of the $\eta$ to
SM particles appear only via EWSB and mixing effects and are therefore
$v/F$ suppressed. The dominant decays of the $\eta$ are to the
heaviest SM fermions (tops, bottoms, and taus), as well as to gluons and
photons via the triangle anomaly. In a model with a heavy $SU(2)$
singlet $T$ quark, the chiral structure of the $\eta$ couplings is
\cite{effective,axions},
\begin{center}
\begin{tabular}{cc}
\renewcommand{\arraystretch}{1.2}
$
  \begin{array}{|cc|}
    \hline
    \bar{T}TH & O(\frac{v}{F}) \\
    \bar{T}tH & O(1)\,\mathcal{P}_L + O(\frac{v}{F})\,\mathcal{P}_R\\
    \bar{t}tH & O(1)\\
    \hline
  \end{array}
$
  & \qquad
$
  \begin{array}{|cc|}
    \hline
    \bar{T}T\eta & O(1)\,\gamma_5 \\
    \bar{T}t\eta & O(1)\,\mathcal{P}_R + O(\frac{v}{F})\,\mathcal{P}_L \\
    \bar{t}t\eta & O(\frac{v}{F})\, \gamma_5\\
    \hline
  \end{array}
$
\end{tabular}
\end{center}
where $\mathcal{P}_{L(R)}$ are the left-(right-)handed projectors,
respectively. 

\section{THE $\mu$ MODEL}

In \cite{axions}, we investigate several different realizations of the
Little Higgs mechanism: the Littlest Higgs, the simple-group model and
the $\mu$ model \cite{little,simplest}. In all of these models the
$\eta$ particle has the gross properties mentioned above. Let us
concentrate here on the $\mu$ model, where the embedding of the $U(1)$
and the quantum numbers of the particles are manifest, and the
predictability is much better than in the other models. The model is a
``moose''-like model, where a $\mu$ term like in the MSSM breaks the
global symmetry explicitly and triggers EWSB. The EW gauge group is
enlarged to $SU(3)_L \times U(1)_Y$ while the global symmetry breaking
is $U(3) \to U(2)$. There are two non-linear sigma model fields,
$\Phi_1 = \exp[ i   F_2/F_1 \Theta] (0,0,F_1)$ and $\Phi_2 = \exp[- i
  F_1/F_2 \Theta] (0,0,F_2)$, with $\Theta$ being the matrix  
\[
  \Theta = \frac{1}{\sqrt{F_1^2 + F_2^2}}
  \begin{pmatrix}
    \begin{array}{cc}
      \eta & 0 \\ 0 & \eta
    \end{array} 
       & h^*   \\
    h^T & \eta
  \end{pmatrix}.
\]
Introducing the parameter $\kappa \equiv F_1/F_2 + F_2/F_1 \geq 2$
the scalar potential can be written as
\begin{equation*}
  -V = - (\delta m^2 + \mu^2 \kappa) (h^\dagger h) - \mu^2 \kappa
   \frac{\eta^2}{2} + (\frac{\mu^2 \kappa^2}{12 F_1F_2} - \delta
   \lambda) (h^\dagger h)^2 + \ldots,
\end{equation*}
where $\delta m^2$ and $\delta \lambda$ are the one-loop contributions
from the Coleman-Weinberg potential. The mass of the pseudo-axion is
then given by $m_\eta = \sqrt{\kappa} \mu \geq \sqrt{2} \mu$, while
the $\eta$ and the Higgs masses are connected by the relation $m_H^2 =
-2 (\delta m^2 + m_\eta^2)$. From this one can read off that $\mu \sim
v$ in order to avoid fine-tuning. But $\mu$ anyhow is
restricted to lie within several GeV and nearly 400 GeV, where the
upper bound comes from the EWSB constraint, the lower bound is due to
the LEP Higgs exclusion limit. The Higgs mass varies between 140 and
800 GeV, depending on the ratio of the two VEVs, $F_1/F_2$. This ratio
plays a role similar to $\tan\beta$ in the MSSM. $m_\eta$ varies
between several GeV and roughly 400 GeV, rising linearly with $\mu$. 

To simplify phenomenology, we assume that there is no mixing between
the SM and heavy fermions for the first two generations, and that the
heavy top mass takes its minimal value. Then the only free parameters
are $F_1$, $F_2$ and $\mu$ with the following bounds: $\sqrt{F_1^2 +
  F_2^2} \gtrsim 2$ TeV from EW precision observables and bounds
on contact interactions, $F_1 \gtrsim v$, $F_2 > F_1$ from fermion 
mixing and the universality of fermion couplings. In
ref.~\cite{simplest}, a so called ``golden point'' is defined for which all
constraints are fulfilled:  $F_1 = 0.5$ TeV, $F_2 = 2$ TeV, $M_T = 1$
TeV, $M_{D,S} = 0.7$ TeV, $M_{W'} = 0.95$ TeV, $M_{Z'} = 1.2$ TeV. $D$
and $S$ are the other heavy quarks (for more details see
\cite{simplest,axions}), and $W',Z'$ are the additional heavy vector
bosons. The pseudo-axion $\eta$ resembles the pseudoscalar
$A$ in a two-Higgs doublet model (2HDM) for small 
\begin{figure}
  \includegraphics[scale=.5]{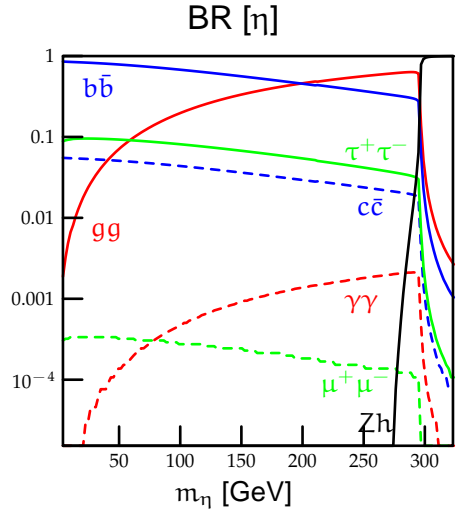}%
  \caption{Branching ratios of the $\eta$ in the $\mu$
  model. \label{fig:branch}} 
\end{figure}
$\tan\beta = F_1/F_2 < 1$. The branching ratios of the $\eta$ for the
golden point are given in fig. \ref{fig:branch}. Concerning the Higgs
phenomenology, there are new decays $H \to Z\eta$, which can amount to
$1-2\%$ BR for a light $\eta$, while $H \to \eta\eta$ is negligible,
but can reach $5-10 \%$ in the extended simple group model.       


\section{COLLIDER SIGNATURES}

Here we discuss the capability of the ILC and the photon collider to
detect such a pseudoscalar particle. Since the largest coupling to SM
\begin{figure}
  \includegraphics[scale=.5]{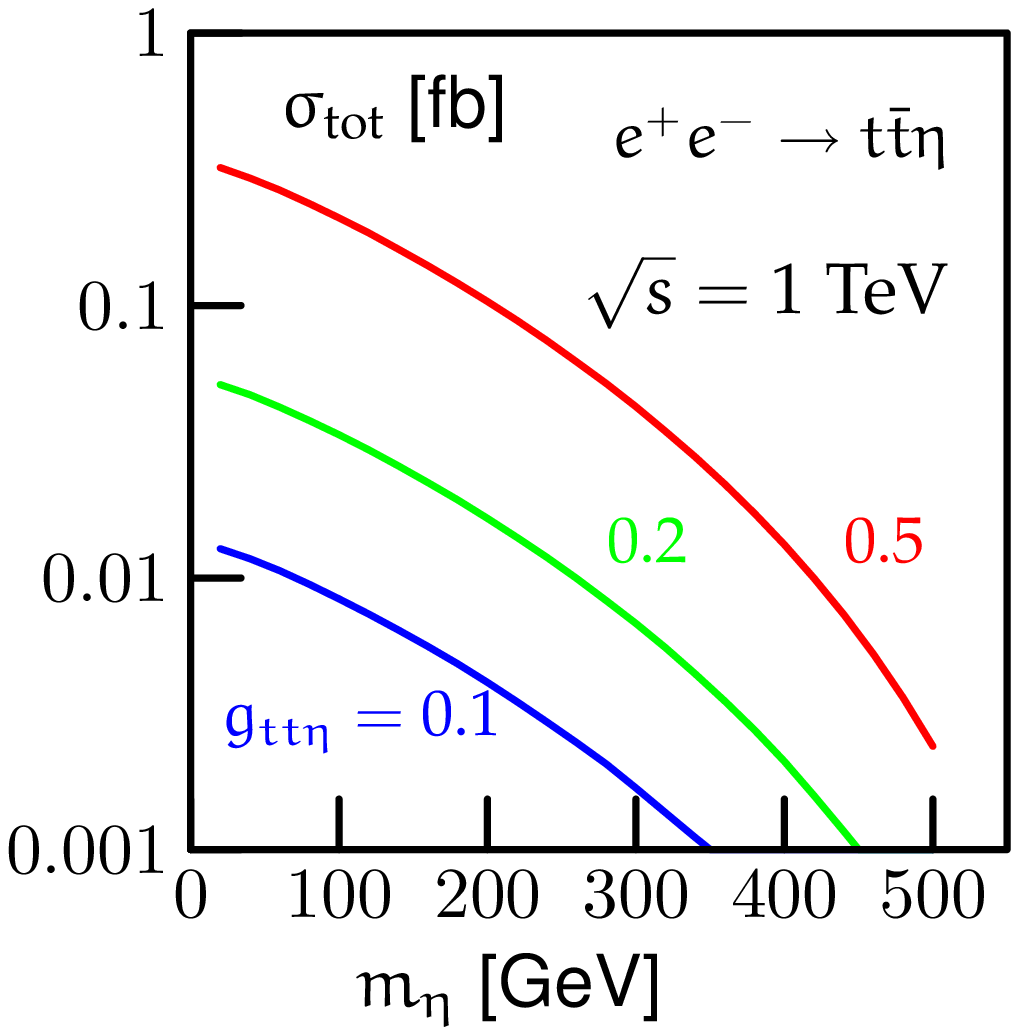}%
  \includegraphics[scale=.5]{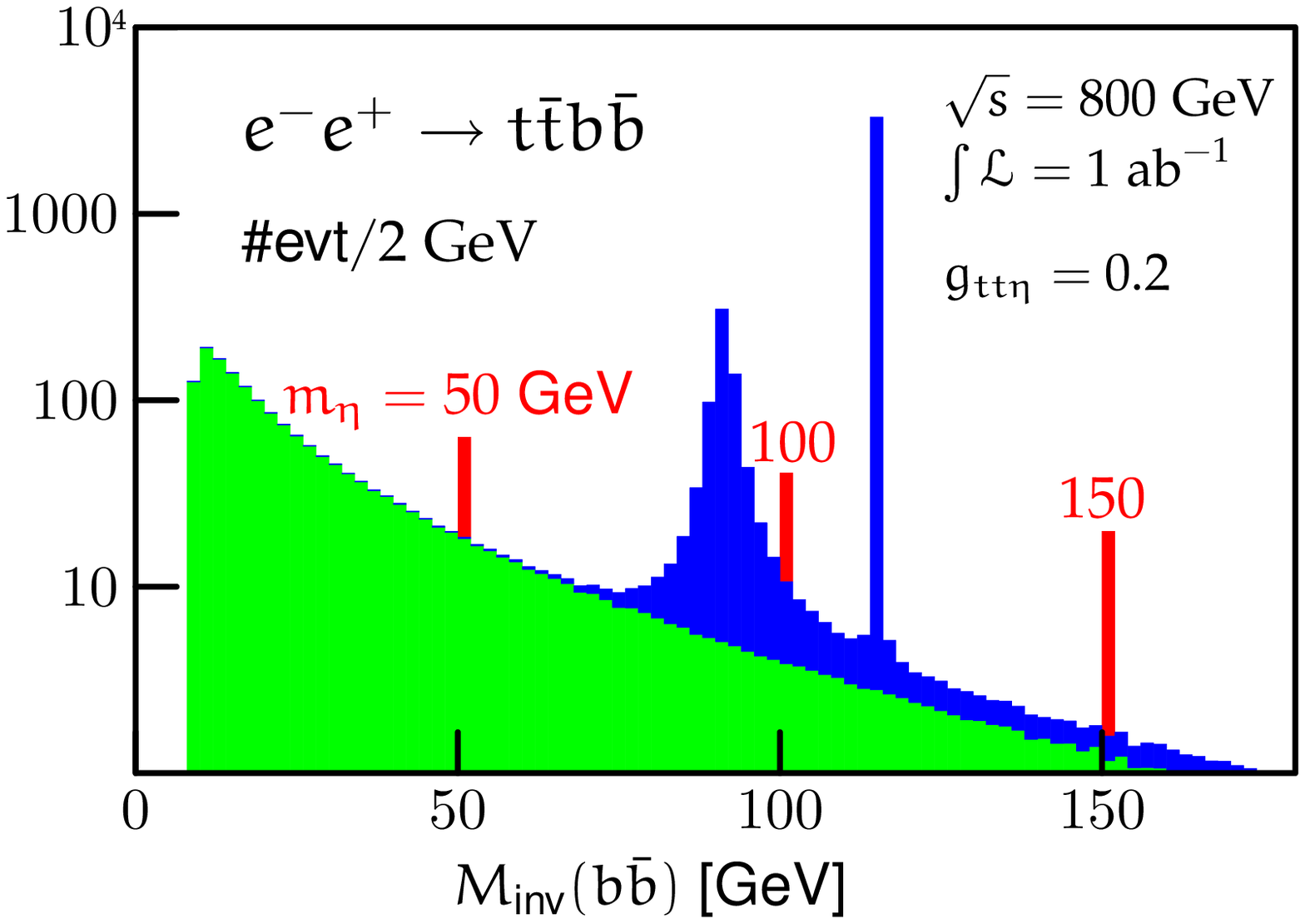}%
  \caption{Associated production of the $\eta$ at the ILC. \label{fig:ilc}}
\end{figure}
particles is the $\eta t\bar t$ coupling, the most promising channel is
associated production with a $t\bar t$ pair and the subsequent decay of the
pseudo-axion into a $b\bar b$ pair. Fig.~\ref{fig:ilc} (left) shows 
the cross section for the associated production depending on the
coupling ratio $g_{tt\eta}/g_{tth}$ and the mass of the $\eta$. The
right plot shows the $b\bar b$ invariant mass spectrum at an $800$ GeV
ILC with high luminosity; the light area is the QCD background, while the dark
area is the EW background, the $Z$ and a 115 GeV Higgs,
respectively. The sharp spikes at $50$, $100$ and $150$ GeV are possible
signals from the $\eta$ resonance. So, except for the case that the
$\eta$ lies too close to the $Z$ or Higgs resonance, it is clearly
visible at the ILC. While for higher $\eta$ masses the cross section
goes down, also the background fades away so that for higher masses
the signal-to-background ratio should be even better.

In the $\mu$ model -- in contrast to the Littlest Higgs model -- there
is a coupling $ZH\eta$, which is $v/F$ suppressed, but 
enhanced by a $\tan\beta$ effect. Therefore it is possible to study
the process $Z^* \to H\eta$ in analogy to $A$ production in the 2HDM
\cite{tocome}. 

At a photon collider which is a precision machine dedicated to Higgs
\begin{figure}
  \includegraphics[scale=.5]{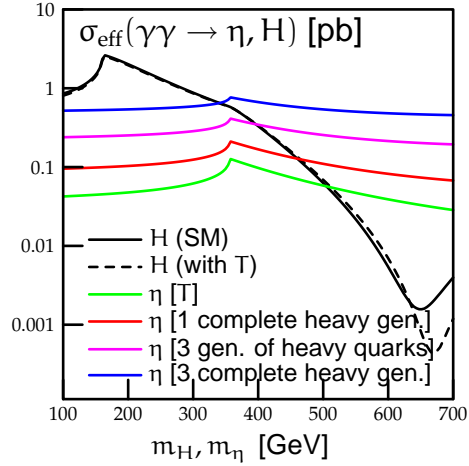}%
  \caption{Effective $\gamma\gamma \to H,\eta$ cross
  sections. \label{fig:gge}} 
\end{figure}
physics, scalar and pseudoscalar particles can be studied as $s$
channel resonances coupled to two photons by their anomaly
couplings. The signal/background ratio is similar to the linear
collider, with the signal cross section going down with increasing
pseudoscalar mass, while the background also vanishes for higher
energies. Fig. \ref{fig:gge} shows the effective production cross
section for the pseudo-axion in Little Higgs models in comparison to
the  SM Higgs production cross section.  For higher masses (over 300
GeV), the cross sections for the pseudoscalar reach the same order of
magnitude as for the SM Higgs, and can be even larger since for the
$\eta$ there is no destructive interference with gauge boson loops. The
cross section depends crucially on the number of particles running in
the loop. From bottom to top, the curves correspond to one heavy top
quark (as in the Littlest Higgs), one complete heavy generation, three
generations of heavy quarks, and three complete heavy generations (as
in the simple-group model), respectively. The spike in the cross
section comes from the interference of the heavy particle loops with
the top loop which is important since the $\eta$ couples  
\begin{figure}
  \includegraphics[scale=.5]{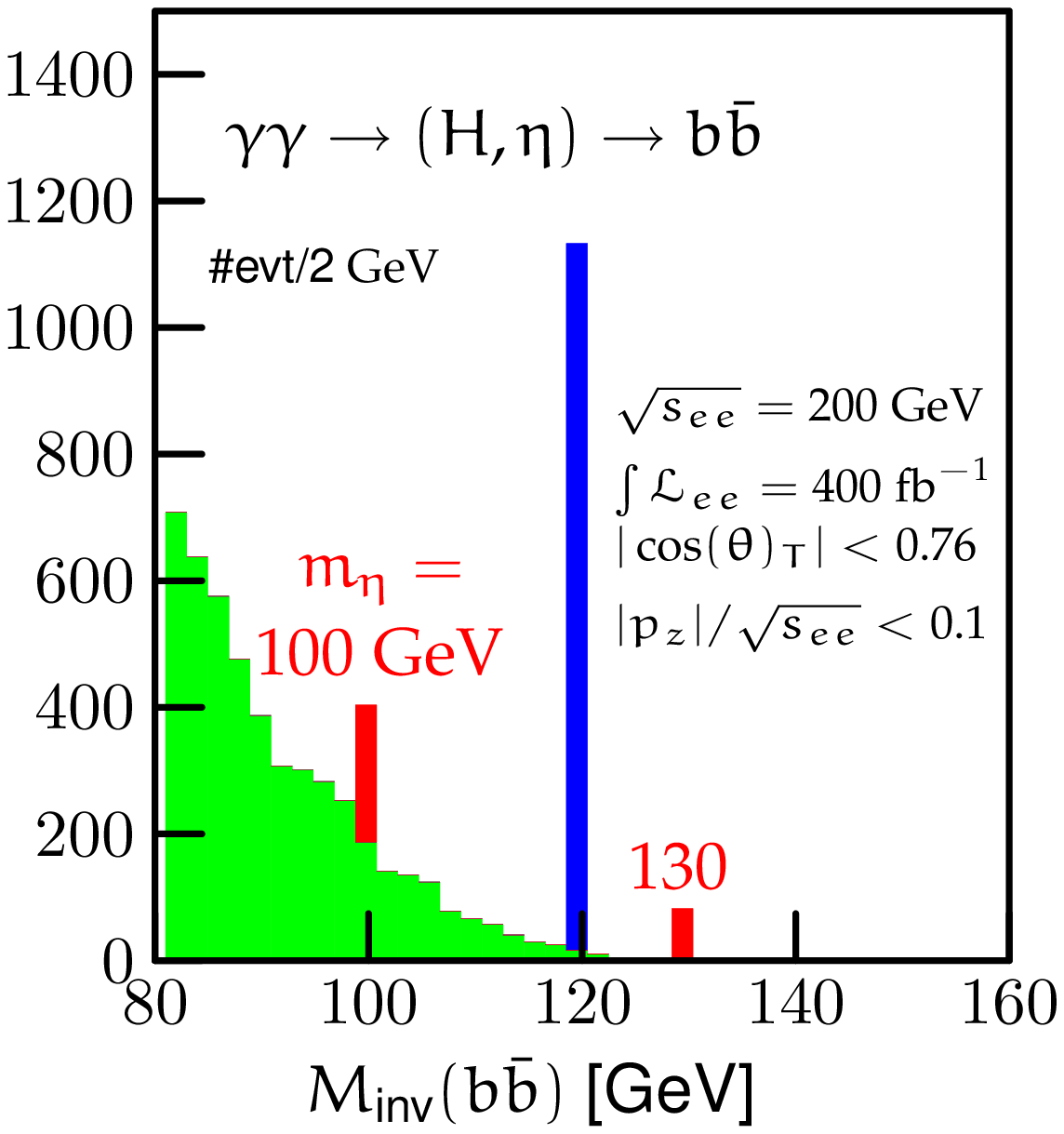}%
  \includegraphics[scale=.5]{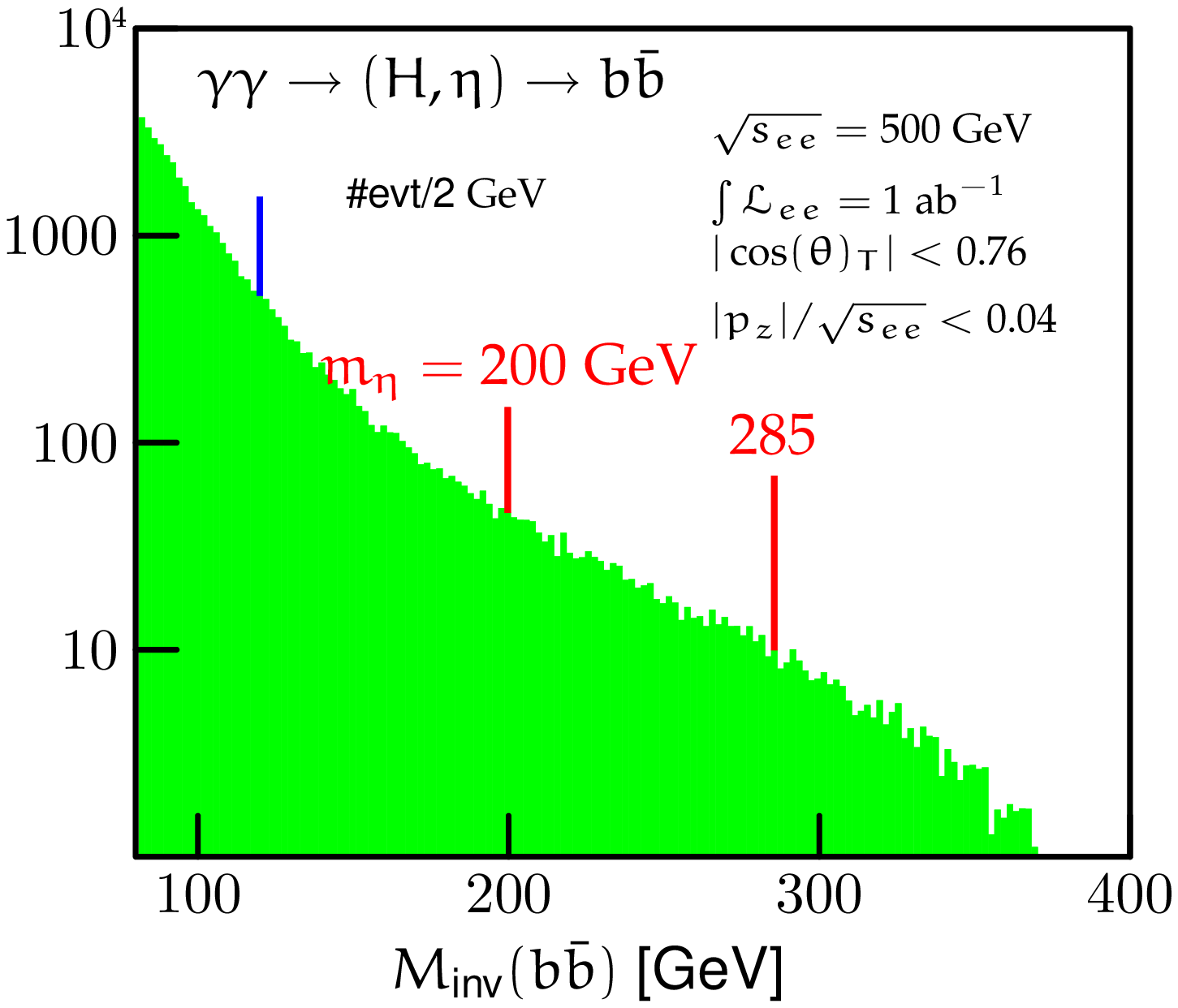}%
  \caption{$\eta$ signal at the 200 and 500 GeV photon
  collider. \label{fig:ggcoll}} 
\end{figure}
with order one to the SM top. 

The collider phenomenology of the $\eta$ at the golden point ($m_\eta
= 310$ GeV) is almost identical to the MSSM pseudoscalar higgs $A$, as
investigated in \cite{maggie}. For the coupling of the $\eta$ to
$b\bar b$
we took $0.45\times g_{bbh}$. The partial width 
$\Gamma(\eta\to\gamma\gamma)$ rises with increasing $m_\eta$ from
$0.15$ keV for $m_\eta = 100$ GeV to $3.6$ keV for $m_\eta = 285$ GeV,
where the last value has been chosen to be just below the threshold
for $\eta \to Zh$ decay, and to stay in the area of a dominant
$b\bar b$ final state. Fig. \ref{fig:ggcoll} shows the $b\bar b$
invariant mass spectra for a photon collider running at 200 and 500
GeV, respectively. The plots have been made using the programs in 
\cite{progs}. Standard cuts have been applied, and a $b$ tagging
efficiency of $80 \%$ has been taken into account. The signal of a
$120$ GeV SM Higgs has been shown for comparison. One can see that
over a wide range of $\eta$ masses a discovery at the photon collider
is possible. To optimize the search, the energy of the linear collider
can be tuned once a significant deviation from the SM spectrum has
been seen \cite{maggie}; for conservative reasons, this was not done
here. For more details, see \cite{axions}. 


\section{CONCLUSIONS}

Little Higgs models are an elegant alternative to supersymmetry to
solve (part of) the hierarchy problem and provide a consistent
framework for EWSB. If there are (approximate) $U(1)_\eta$
symmetries in the model, this gives rise to new EW singlet
pseudoscalars, called pseudo-axions of Little Higgs models. In some
models, these particles are absorbed as longitudinal modes
of heavy $Z'$ states, which would generally be easy to detect at future
colliders. If these particles are in the physical spectrum, they
would be difficult to detect, but alter the phenomenology of the Higgs
and the new heavy quark states present in this class of models. An
explicit breaking of the $U(1)$ symmetry by Yukawa and gauge
interactions circumvents the axion limits from astrophysics. At the
LHC the pseudoscalar could be detected in gluon fusion as a signal in
the diphoton spectrum, if it is heavy enough. For smaller $\eta$ masses
the linear and photon colliders would be well-suited to detect
such a state, where the linear collider could specially cover the low
$\eta$ mass range, and would give a spectacular opportunity (in some of the
models) via the $Z^* \to H\eta$ process. The photon collider could 
search for the pseudo-axion over a wide range of masses, enabling a
measurement complementary to the linear collider and also to possibly
give access to branching ratios of the $\eta$. 


\begin{acknowledgments}
This research was supported in part by the National Science Foundation
under Grant No. PHY99-07949, the U.S. Department of Energy under grant
No. DE-FG02-91ER40685, and by the Helmholtz-Gemeinschaft under Grant
No. VH-NG-005.
\end{acknowledgments}



\begin{thebibliography}{9}   

\bibitem{little}
  N.~Arkani-Hamed, A.~G.~Cohen, H.~Georgi,
  Phys.\ Lett.\ B {\bf 513} (2001) 232;
  N.~Arkani-Hamed, A.~G.~Cohen, T.~Gregoire, and J.~G.~Wacker,
  JHEP\ {\bf 0208} (2002) 020.
  M.~Schmaltz, Ann.\ Rev.\ Nucl.\ Part.\ Sci.\
  arXiv:hep-ph/0502182.
\bibitem{simplest}
  D.~E.~Kaplan, M.~Schmaltz, JHEP {\bf 0310} (2003) 039;
  M.~Schmaltz, JHEP {\bf 0408} (2004) 056. 
\bibitem{effective}
  W.~Kilian, J.~Reuter, Phys.\ Rev.\ {\bf D70} (2004) 015004.
\bibitem{axions}
  W.~Kilian, D.~Rainwater, J.~Reuter, Phys.\ Rev.\ {\bf D 71} (2005), 015008. 
\bibitem{tocome}
  W.~Kilian, D.~Rainwater, J.~Reuter, in preparation.
\bibitem{pdg}
  S. Eidelman et al., Phys.\ Lett.\ {\bf B592} (2004), 1. 
\bibitem{maggie}
 M.~M.~M\"uhlleitner, M.~Kr\"amer, M.~Spira and P.~M.~Zerwas,
  Phys.\ Lett.\ B {\bf 508}, 311 (2001).
\bibitem{progs}
  T. Ohl, arXiv:hep-ph/0011243;
  M.~Moretti, T.~Ohl, J.~Reuter,
  arXiv:hep-ph/0102195;
  T.~Ohl, Comput.{} Phys.{} Commun.{} \textbf{101} (1997) 269;
  T.~Ohl, WUE-ITP-2002-006;
  W.~Kilian, LC-TOOL-2001-039, Jan 2001.
\end{thebibliography}
\end{document}